\documentclass[aps,prl,preprint,showpacs]{revtex4}
\usepackage{mathrsfs}
\usepackage{graphicx}
\usepackage{amssymb}
\usepackage{amsmath}

\usepackage{setspace}

\begin{document}

\title{Single-Photon Imaging and Efficient Coupling to Single Plasmons}

\author{M. Celebrano, R. Lettow, P. Kukura, M. Agio, A. Renn, S. G\"otzinger, V. Sandoghdar}

\affiliation{Laboratory of Physical Chemistry and optETH, ETH
Zurich, CH-8093 Zurich, Switzerland}

\begin{abstract}

We demonstrate strong coupling of single photons emitted by
individual molecules at cryogenic and ambient conditions to
individual nanoparticles. We provide images obtained both in
transmission and reflection, where an efficiency greater than 55\%
was achieved in converting incident narrow-band photons to
plasmon-polaritons (plasmons) of a silver nanoparticle. Our work
paves the way to spectroscopy and microscopy of nano-objects with
sub-shot noise beams of light and to triggered generation of single
plasmons and electrons in a well-controlled manner.
\end{abstract}


\maketitle

Coupling of light to dipolar radiators lies at the heart of
light-matter interaction. Theoretical studies have predicted more
than 80\% extinction of a focused classical Gaussian beam by a
single dipolar radiator~\cite{Zumofen:08,Mojarad:09}, and recent
experimental investigations have reported up to 12\% extinction of
classical light by single quantum
emitters~\cite{Vamivakas:07,Wrigge:08,Tey:08}. Laboratory
realizations of spectroscopy and microscopy on single nano-objects
with single-photon illumination, however, have been confronted by
obstacles. In particular, excitation of a quantum emitter by
individual photons has only been possible in confined
geometries~\cite{Eschner:01,Rempe:03}, and quantum optical
imaging~\cite{Kolobov:99,Treps:02,Lugiato:02} of subwavelength
structures has not been explored at the single-photon level. One
bottle-neck in spectroscopy with single photons is access to bright,
tunable, and narrow-band single-photon sources~\cite{Lounis:05}.
Another challenge stems from the fundamental wave character of
propagating photons which leads to a weak coupling with matter. In
this Letter, we demonstrate more than 55\% coupling between a
diffraction-limited beam of single photons and single silver
nanoparticles, which act as classical dipolar antennae. This strong
photon-dipole coupling allows efficient excitation of single
plasmon-polaritons (plasmons)~\cite{Akimov:07,Tame:08} and imaging
of nano-objects with nonclassical light.

\begin{figure} [b]
\begin{center}
\includegraphics[width=7cm]{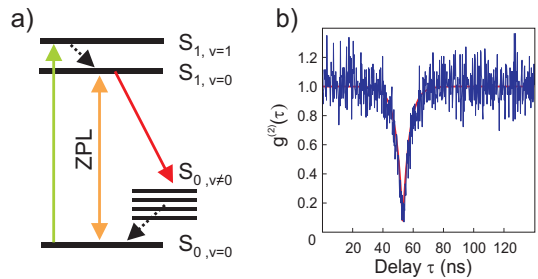}
\caption{ (a) The energy level scheme of a dye molecule. See text
for details. (b) An example of the second-order correlation function
of a single DBATT molecule under continuous-wave excitation.
\label{SPS}}
\end{center}
\end{figure}

The source of single photons in our experiments is a single dye
molecule embedded in an organic crystalline matrix.
Figure~\ref{SPS}a shows the energy level scheme of such a molecule
as well as its excitation and fluorescence channels. We begin with
experiments performed at $T\simeq1.4$~K, where we used a tunable
narrow-band dye laser to excite dibenzanthanthrene (DBATT) embedded
in n-tetradecane on the $S_{0, \rm v=0}\rightarrow S_{1, \rm v=1}$
transition at the wavelength of
$\lambda\simeq581$~nm~\cite{Lettow:07}. This state rapidly decays to
the $S_{1, \rm v=0}$ state that has a lifetime of 9.5~ns determined
by fluorescence to the $S_{0, \rm v}$ states. By filtering the broad
Stokes-shifted fluorescence to the $\rm v\neq0$ manifold and
collecting the emission on the $S_{1, \rm v=0}\rightarrow S_{0, \rm
v=0}$ zero-phonon line (ZPL), we obtained a source of single photons
at $\lambda\simeq589$~nm with a lifetime-limited linewidth of
17~MHz~\cite{Lettow:07}. Figure~\ref{SPS}b displays a recorded
second-order autocorrelation function that proves the strongly
photon-antibunched nature of this light. More details of the
cryogenic setup and characterization of the single-photon source can
be found in Ref.~\cite{Lettow:07}. Here, it suffices to point out
that this narrow-band single-photon source delivers a high power of
up to 500~fW, corresponding to about $10^6$ detected photons per
second.

\begin{figure}[b]
\begin{center}
\includegraphics[width=8cm]{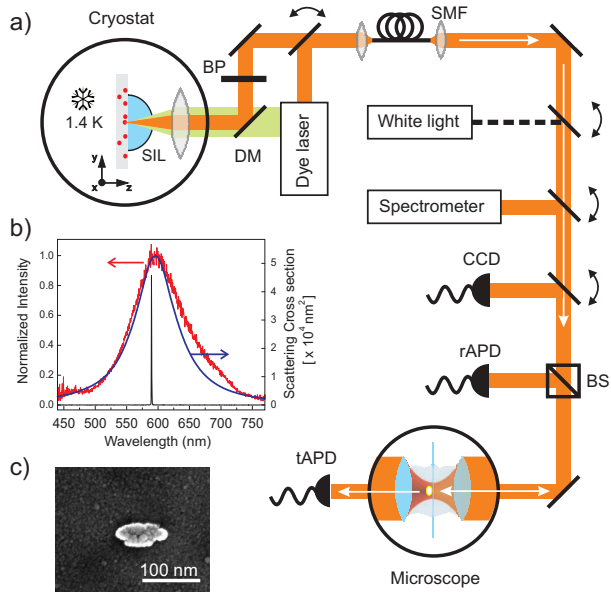}
\caption{(a) Single dye molecules embedded in a thin organic matrix
at T=1.4~K produce a beam of single photons. This beam is collected
and collimated by a solid-immersion lens and an aspherical lens
inside the cryostat and then coupled into a single-mode fiber (SMF).
The output of this fiber is sent to the sample in a room-temperature
microscope. Two avalanche photodiodes rAPD and tAPD register the
signal in reflection and transmission, respectively. A spectrometer
records the plasmon spectrum of a particle upon illumination by a
white-light source. DM: dichroic mirror, BS: beam splitter. (b) The
red curve shows the experimentally measured plasmon spectrum of the
particle studied in the first experiment. The blue curve displays a
theoretical spectrum corresponding to an ellipsoidal silver particle
with long and short axes of 94 and 46~nm, respectively. The black
curve shows the spectrum of the narrow-band single-photon source.
(c) Electron microscope image of a typical silver spheroid.
\label{setup}}
\end{center}
\end{figure}

The collimated beam of single photons was coupled into a single-mode
fiber and directed to a home-built microscope at room temperature as
shown in Fig.~\ref{setup}a. An oil-immersion objective with a
numerical aperture (NA) of 1.4 focused this light onto single silver
nanoparticles with nominal diameter of 60~nm (British Biocell),
which were spin coated on a glass cover slide and index-matched by
immersion in oil (refractive index=1.49). Another microscope
objective (NA=1.4) collected the transmitted light and sent it onto
an avalanche photodiode (APD). A second APD was used to record the
signal in reflection. In addition, we used flip mirrors to couple
the light from the dye laser or a white-light source directly to the
room-temperature microscope for characterization and diagnostics of
the nanoparticles on the sample. By inserting a pinhole in the image
plane, we could select each single particle and record its plasmon
spectrum using a grating spectrometer.

The red trace in Fig.~\ref{setup}b plots the plasmon spectrum of a
nanoparticle that matched our single photon source at
$\lambda=589$~nm indicated by the black curve. This spectrum
corresponds to a prolate silver ellipsoid with a short axis of 46 nm
and a long axis of 94 nm that is parallel to the substrate. We find
good agreement with the results of calculations that considered a
dipolar scatterer~\cite{EPAPS} and illumination parallel to the long
axis of the particle (see blue curve of Fig.~\ref{setup}b). Indeed,
electron microscopy revealed that the colloidal particles were
mostly elongated (see Fig.~\ref{setup}c) with a notable variation in
shape and size. We, thus, selected nanoparticles that matched the
wavelength of our single-photon source (see Fig.~\ref{setup}b) and
maximized the signal.

\begin{figure}[b]
\begin{center}
\includegraphics[width=8cm]{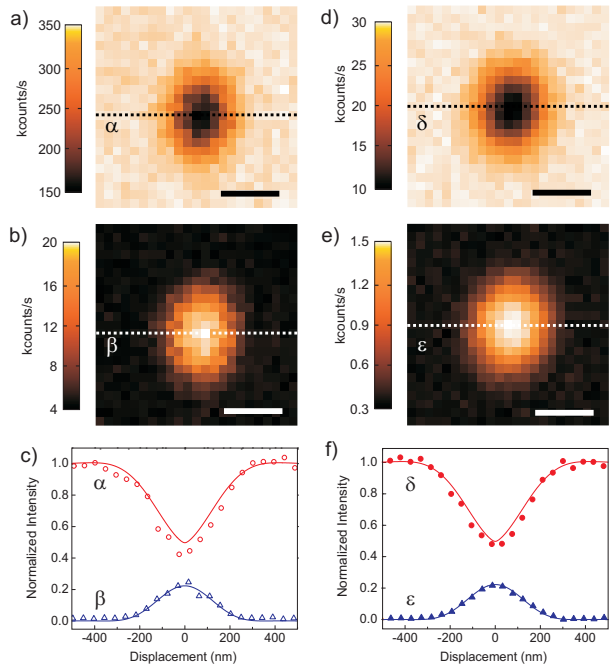}
\caption{Transmission (a) and reflection (b) images obtained when
the sample was scanned laterally across the focus of a laser beam at
a speed of 10 ms per pixel. (c) Cross sections from (a) and (b).
Average of 12 transmission (d) and reflection (e) images obtained
when the sample was laterally scanned in the focus of the
single-photon beam at 40 ms per pixel. (f) Cross sections from (d)
and (e). Light beams were polarized along the vertical directions of
the images in (a), (b), (d), and (e). Scale bars correspond to
500~nm. \label{SP-zooms}}
\end{center}
\end{figure}

Figures~\ref{SP-zooms}a and b display images of a nanoparticle
recorded simultaneously in transmission and reflection as the sample
was scanned by a piezo-electric stage across the focus of the laser
beam. The origin of these signals is scattering of the incident
light from the nanoparticle~\cite{Bohren-83book}, and the details of
the contrast mechanism and the detection scheme are discussed in the
literature~\cite{Mikhailovsky:03,Lindfors:04}. Here, we briefly
highlight the interferometric character of the signal $I_{\rm d}$
recorded on the detector, which is described as
\begin{equation}
I_{\rm d}=\left\vert E_{\rm ref}+E_{\rm sca}\right\vert
^{2}=\left\vert E_{\rm ref}\right\vert ^{2}+ \left\vert E_{\rm
sca}\right\vert^{2}-2\left\vert E_{\rm ref}\right\vert\left\vert
E_{\rm sca}\right\vert \sin \varphi. \label{signal}
\end{equation} where $E_{\rm sca}$ is the electric field of the scattered light at
the detector position, and $E_{\rm ref}$ denotes the electric field
of a ``reference" beam. In case of the transmission signal, the
reference is the incident light, whereas for the reflection signal,
it is produced by a residual reflection of the illumination within
the optical setup. Depending on the relation between $E_{\rm ref}$
and $E_{\rm sca}$, the second or the third term of Eq.~(1) might
dominate and determine the signal contrast~\cite{Lindfors:04}.
Furthermore, this is influenced by the scattering phase angle
$\varphi$, which depends on the dielectric function of the
nanoparticle at the illumination wavelength as well as its size and
shape~\cite{Lindfors:04}.

Next, in Figs.~\ref{SP-zooms}d and e we present raster-scan images
of the same single nanoparticle recorded under illumination by
single photons. As shown by the two cross sections $\delta$ and
$\epsilon$ in Fig.~\ref{SP-zooms}f, we find full width at
half-maximum (FWHM) values of 300~nm and 260~nm in transmission and
reflection images, respectively. The solid curves display the
outcome of rigorous vectorial three-dimensional
calculations~\cite{EPAPS} considering a focused Gaussian
beam~\cite{Mojarad:09}. Here, the silver nanoparticle was modeled as
a dipolar emitter, taking into account radiative and dynamic
depolarization corrections~\cite{Meier:83}. We found a good
agreement between the theoretical predictions and the experimental
results if we considered the polarizability corresponding to the
scattering cross section reported in Fig.~\ref{setup}b~\cite{EPAPS}.
A correction to the observed FWHM, $\Delta X_{np}$, for the finite
size of the nanoparticle could also be made according to $\Delta
X_{np}\approx\sqrt{(\Delta X_{pd})^2 + D^2_{np}}$ where
$D_{np}=46$~nm is the particle size along the scan direction, and
$\Delta X_{pd}$ is the FWHM calculated for a point dipole. However,
this amounts to an adjustment of only 3~nm, which we have chosen to
neglect here. We mention in passing that the appearance of the
elliptical images in Figs.~\ref{SP-zooms}a,b,d,e is a well-known
effect for tightly focused linearly-polarized
light~\cite{Richards1959}. Another noteworthy point is that because
of the nearly index-matched sample, the large cross section of the
silver particle, and a tight focusing, the reflection signal is
dominated by the second term of equation (1) and, therefore, maps
the intensity of the incident beam in the focus spot. The
transmission signal, on the other hand, has a substantial
contribution from the interference term that depends on the electric
field of the excitation light, which has a larger spatial extent
than the intensity~\cite{EPAPS}. Indeed a comparison of the data in
Figs.~\ref{SP-zooms}c and ~\ref{SP-zooms}f reveals that the FWHMs in
transmission are systematically wider than those in reflection
images.

Tight focusing is key to achieving a strong coupling between a light
field and a dipolar emitter~\cite{Zumofen:08,Mojarad:09,Wrigge:08}.
The cross sections in Fig.~\ref{SP-zooms}f reveal large extinction
and reflection contrasts of 55\% in transmission and 22\% in
reflection. This is in very good agreement with the rigorous
vectorial calculations shown by the solid curves, which take into
account the modal character of a Gaussian beam as well as the
illumination and collection solid angles~\cite{Zumofen:08}.

The data presented above clearly show the large effect of a single
dipolar oscillator on a propagating light beam. Given that the
interaction of the incoming photons with the nanoparticle has been
mediated by the excitation of its plasmon-polaritons, these results
indicate a high probability that an incident photon excites a single
plasmon~\cite{Tame:08}. To define an efficiency for the conversion
of a photon to a plasmon, we add the probabilities that it is
absorbed and scattered by the particle. In general, computation of
this quantity from the reflection and transmission measurements
requires a careful consideration of the incident focusing angle
$\alpha$ and the collection angle
$\beta$~\cite{Zumofen:08,Mojarad:09}. However, if $\alpha=\beta$, a
simple argument based on energy conservation lets us conclude that
the sum of the scattered and absorbed powers equals the power
removed from the incident beam, which is directly read from the
transmission dip. If $\alpha<\beta$, some of the light that is
scattered at larger angles is also collected, which reduces the
transmission dip. In our case, the illumination and collection
microscope objectives were identical, but the former was not
completely filled in order to minimize the loss of photons. Thus,
the data in Fig.~\ref{SP-zooms}f yield a lower bound of 55\% for the
photon-plasmon conversion efficiency.

The two series of images in Fig.~\ref{SP-zooms} acquired with laser
light and a single photon source appear nearly identical. However,
in the first case the signal can be described by the interference of
classical fields, while the contrast mechanism of the images
recorded by the latter can only be understood as the result of a
Young-double-slit type of experiment for single
photons~\cite{Taylor:09,Davis:88}. Here, the two interfering paths
for each photon correspond to scattering by the nanoparticle and
transmission without any interaction. After averaging the signal
accumulated from a large number of single photons at each pixel, one
retrieves the results familiar from classical optics.

We have demonstrated that a focused beam of single photons can be
used to detect and image nanoparticles. This beam can also be
produced in a triggered fashion by using a pulsed excitation of the
molecule on the $S_{0, \rm v=0}\rightarrow S_{1, \rm v=1}$
transition~\cite{Ahtee:09}. If the excitation beam is strong enough,
one can ensure the production of a photon after each pulse, yielding
an intensity-squeezed light source with a well-defined number of
photons per unit time. Such a light source would allow the detection
of objects with arbitrarily small optical contrast because it
eliminates noise on the first term $\left\vert E_{\rm
ref}\right\vert ^{2}$ of Eq.~(1) so that the second and third terms
can be deciphered regardless of their
magnitudes~\cite{Lindfors:04,Lounis:05,Wrigge:08b}. One should bear
in mind, however, that any loss in the optical system reduces the
degree of squeezing~\cite{Kolobov:99}. In our experiment, the
central source of loss has been the limited collection angle of the
lens used behind the solid-immersion lens in the cryostat (see
Fig.~\ref{setup}a). This can be substantially improved by employing
different choices of lenses~\cite{Koyama:99}.

\begin{figure}[b]
\begin{center}
\includegraphics[width=7cm]{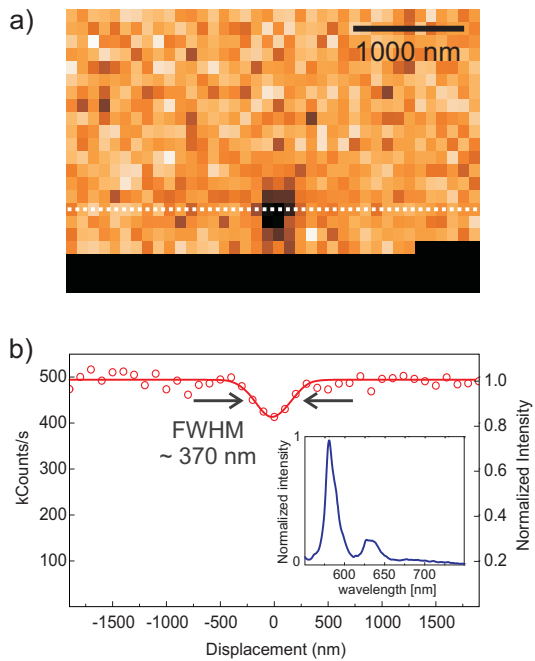}
\caption{a) A raster-scan image of a silver nanoparticle illuminated
by single photons from a terrylene molecule at room temperature. The
black region at the bottom indicates loss of signal caused by the
photobleaching of the molecule. b) A cross section from part (a).
The inset displays the emission spectrum of a single terrylene
molecule. \label{room-temp}}
\end{center}
\end{figure}

A second major cause for losses in the setup of Fig.~\ref{setup}a is
the coupling into an optical fiber. To verify that use of a single
mode and spatial mode filtering is not a strict requirement for the
ability to focus single photons to the diffraction limit, we also
performed free-beam measurements. Here, we collimated the single
photon emission of a room-temperature terrylene molecule embedded in
a thin para-terphenyl film~\cite{pfab:04,Lounis:00}, and sent it
directly to the second microscope as described earlier in
Fig.~\ref{setup}a. Figure~\ref{room-temp}a displays an example of an
image of a silver nanoparticle recorded in this fashion, while
Fig.~\ref{room-temp}b shows a cross section from it. We find that
the full width at half-maximum is as small as 370~nm~\cite{note},
demonstrating that freely propagating single photons can be indeed
focused tightly. Here, the transmission dip amounts to only 15\%
because in this experiment we did not search for a nanoparticle with
a plasmon resonance that matched the emission spectrum of terrylene.
Moreover, the photon-plasmon coupling efficiency in this arrangement
is less efficient than the narrow-band single photon source
discussed earlier because the broad room-temperature emission of
terrylene (see inset in Fig.~\ref{room-temp}b) does not fully
overlap with the particle plasmon resonance.

Strong focusing of single photons demonstrated in this work opens
doors to many interesting experiments where photons are to be
managed with high efficiency and funneled to other quantum systems
in the condensed phase. For example, one can use a silver
nanoparticle as a nanoantenna~\cite{Knight:07} to convert single
propagating photons to single plasmons in nano-circuits. As opposed
to the near-field coupling of photons from pre-positioned emitters
to nanowires~\cite{Akimov:07}, coupling via propagating beams has
the great advantage of being versatile with potential for broad-band
communication because a large number of narrow-band single photon
beams can be coupled simultaneously or sequentially via the same
nano-antenna port. Plasmons can in turn generate electrons in a
photovoltaic process where a quantum of excitation at optical
frequencies gives birth to an electron~\cite{Falk:09}. These
processes would offer interesting possibilities for quantum state
engineering of hybrid systems. Another important promise of our work
is in the detection and imaging of very small nanoparticles and
single molecules~\cite{Lindfors:04,Lounis:05,Kukura:09}. To achieve
this, we plan to optimize the use of solid-immersion lenses to reach
collection efficiencies in excess of 90\%~\cite{Koyama:99}, and thus
produce an intensity-squeezed train of single photons from a single
molecule.

We thank J. Hwang; N. Mojarad, and G. Zumofen for fruitful
discussions. This work was funded by the Swiss National Science
Foundation (SNF).


\begin{thebibliography}{28}
\bibitem{note}
\bibinfo{note}{This value is larger than that expected from a simple estimate of the diffraction
limit because of 1) the inherently larger FWHM that is obtained in
transmission measurements and 2) the lower effective numerical
aperture due a smaller beam diameter in this experiment.}
\end{thebibliography}
\end{document}